\documentclass[aps,pra,a4paper,reprint,amsmath,amssymb,graphicx,longbibliography, showkeys, superscriptaddress]{revtex4-1}

\usepackage[english]{babel}
\usepackage[utf8]{inputenc}
\usepackage[colorinlistoftodos, color=green!40, prependcaption]{todonotes}
\usepackage{float} 
\usepackage{xcolor}

\usepackage[unicode]{hyperref}
\hypersetup{
   unicode=true,          
   plainpages=false,
   colorlinks=true,       
   citecolor=blue,        
}
{}

\usepackage[T1]{fontenc}

\DeclareMathOperator{\cosec}{cosec}

\newcommand{\lJ}{\xi_\mathrm{J}}

\begin{document}

\title{Distributed vorticity model for vortex molecule dynamics
}
\author{Sarthak Choudhury}
\affiliation{Dodd-Walls Centre for Photonic and Quantum Technologies, New Zealand Institute for Advanced Study, Centre for Theoretical
Chemistry and Physics, Massey University, Auckland 0632, New Zealand}

\author{Joachim Brand}
\affiliation{Dodd-Walls Centre for Photonic and Quantum Technologies, New Zealand Institute for Advanced Study, Centre for Theoretical
Chemistry and Physics, Massey University, Auckland 0632, New Zealand}
\date{\today} 

\begin{abstract}
    We analyze the effect of a hard wall trapping potential on the dynamics of a vortex molecule 
    in a two-component Bose-Einstein condensate with linear coherent coupling.
    A vortex molecule consists of a vortex of the same charge in each component condensate connected by a domain wall of the relative phase.
    In a previous paper [\href{https://link.aps.org/doi/10.1103/PhysRevA.106.043319}{Phys.\ Rev.\ A 106, 043319}] we described the interaction of a vortex molecule with the boundary using the method of images by separately treating each component vortex as a point vortex, in addition to a Magnus force effect from the surface tension of the domain wall. Here we extend the model by considering a continuous distribution  of image vorticity reflecting the effect of the domain wall on the vortex molecule phase structure.
    In the case of a precessing centered vortex molecule in an isotropic trap, distributing the image vorticity weakens its contribution to the precession frequency.
    We test the model predictions against numerical simulations of the coupled Gross-Pitaevskii equations in a two-dimensional circular disc and find support for the improved model.
\end{abstract}

\keywords{coherently coupled superfluids, fractional vortex molecule, coreless vortex, Josephson vortex}

\maketitle


\section{Introduction} 
    \label{sec:outline}
    The quantization of vortex lines is a striking feature of superfluids
    that appears as a consequence of 
    Bose-Einstein condensation \cite{Leggett2006}. The dynamics of superfluid vortices 
    still poses many open questions and is actively pursued
    \cite{Yu2017,Johnstone2019,kwonSoundEmissionAnnihilations2021,Caracanhas2021}.
    Shortly after the first observation of quantized vortices in a dilute gas Bose-Einstein condensate (BEC) \cite{Anderson1999}, experiments introduced coherent coupling in 
    multi-component BECs by applying a radio-frequency electromagnetic field that drives a Rabi transition between internal (hyperfine) states in the constituent atomic gas \cite{Matthews1999a},
    with more refined experiments becoming available recently \cite{Nicklas2015,Farolfi2021,Farolfi2021a}. Theoretical work then examined the peculiar structure of vortices in such a two-component BEC under the continuous influence of coherent coupling \cite{Son2002,Garcia-Ripoll2002}. Due to the coherent coupling, the phases of the component condensates align in equilibrium. As a consequence, vortices in the component BECs have to be connected by a vortex line that extends between the two 
    BECs as a domain wall of the relative phase, also known as a Josephson vortex. Analytic solutions for stationary Josephson vortices were first found by Kaurov and Kuklov \cite{Kaurov2005,Kaurov2006}, and families of moving Josephson vortices were characterized in Ref.~\cite{Shamailov2018}. 
    
    Domain walls have an energy content, or surface tension, which is approximately linear in their extent, i.e.~length in two dimensions and area in three dimensions. This makes them susceptible to breaking up into smaller fragments. Their dynamical stability is determined by the sign of their effective mass, which depends on the competition between the Rabi coupling and the nonlinear mean-field energy in the BECs \cite{Shamailov2018,Ihara2019,Gallemi2019}.
    Interesting analogies to axions and quark confinement were first pointed out by Son and Stephanov \cite{Son2002} (see also \cite{etoCollisionDynamicsReactions2020,Yasui2020,Kobayashi2022}). 

    In two-dimensional BECs, a domain wall can either terminate at a boundary of the superfluid domain, or at an appropriately charged vortex in either of the component BECs. A configuration with same-charge vortices in each component connected by a domain wall is known as a vortex molecule \cite{Kasamatsu2004}. Sometimes it is referred to as fractional vortex molecule to highlight the fact that the quantized vortex charge can be thought of as being split into fractional charges residing in separate locations at the vortices in each component BEC \cite{Eto2018}. Theoretical studies of equilibrium configurations have been extended to vortex molecule lattices \cite{Cipriani2013}. Topological defects analogous to vortex molecules are being investigated experimentally in superfluid $^3$He \cite{Makinen2019}.

    The dynamics of vortex molecules has first been considered by Tylutki \emph{et al.} \cite{Tylutki2016} in the context of a two-dimensional coupled BEC in an isotropic harmonic trap. In this scenario a symmetrically centered vortex molecule rotates with a constant angular frequency around the trap axis, referred to as precession.
    The vortex molecule dynamics was described by the superposition of three separate velocity components: One derived from the influence of the harmonic trapping potential on the individual (point) vortices in a Thomas Fermi approximation, and two contributions from the Magnus effect related to a short-range core repulsion and an attractive force due to the surface tension of the domain wall.
    A generalised Magnus force on a quantized vortex gives rise to a transverse velocity component according to $\mathbf{F} = 2\pi n_0 \hbar \hat{\kappa}\times \mathbf{V}$, where $n_0$ is the superfluid density, $\hat{\kappa}$ is the circulation unit vector, and $\mathbf{V}$ is the velocity of the vortex relative to that of the background superfluid \cite{Vinen1961,Thouless1996}.
    Calderaro  \emph{et al.} \cite{Calderaro2017a} then developed a Lagrangian variational formalism focussing on the effect of the domain wall on the vortex dynamics. They obtained analytic results in two different regimes: The attractive Magnus force is linear in the molecular distance $d$ (the length of the domain wall) in the regime of weak Rabi coupling where
    $\xi_\mathrm{J} \gg d \gg \xi$, and $\xi_\mathrm{J}$ is the Josephson vortex length scale (width of the domain wall), and $\xi$ is the condensate healing length. In this regime, the Magnus force contribution to the precession frequency is constant. The other regime of strong Rabi coupling where  $d \gg \xi_\mathrm{J} \gg \xi$ is the one considered by Tylutki \emph{et al.} \cite{Tylutki2016} where the Magnus force is constant and provided by the surface tension of an infinite domain wall. 
    
    In a previous work we developed an extended point vortex model to analyze the dynamics of a vortex molecule in a flat-bottom trap realizing a channel geometry with parallel hard walls \cite{Choudhury2022}. The model includes the Magnus-force effects of the domain wall and possible core repulsion by parameterizing a numerically-obtained vortex-vortex interaction energy in the absence of domain boundaries. The
    effects of the hard-wall boundaries are separately taken into account by the method of images applied on the individual component vortices. This theory was able to predict all the qualitative features in the pendulum-like phase space of the vortex molecule dynamics in the channel geometry of Ref.~\cite{Choudhury2022}.

    In this work we aim for a more accurate description of the vortex-molecule dynamics in the presence of hard-wall trapping potentials. 
    Instead of treating the component vortices as localized point vortices, we consider the vorticity to be distributed along the domain wall connecting the component vortices for the purpose of generating image vorticity. This is motivated by the fact that over length scales larger than the molecular separation and the Josephson length scale, the phase of both condensate components aligns in numerical simulations.

\begin{figure}
        \centering
        \includegraphics[width=1\linewidth]{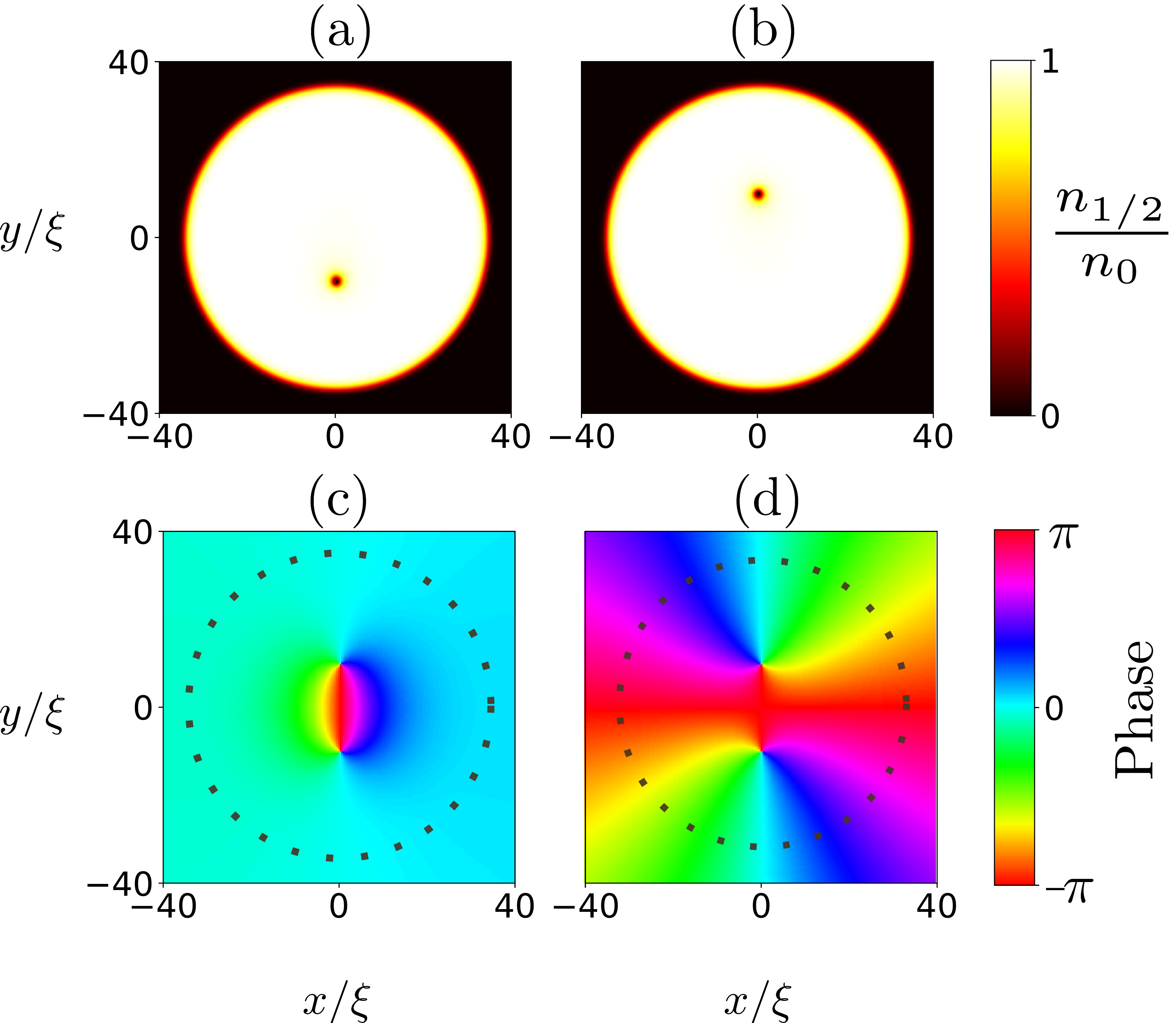}
        \caption{Vortex molecule in a disc. Numerical solution of the GPE \eqref{coupledGPE} with a centered vortex molecule with molecular length $d=20\xi$ in a disc-like trap described by Eq.~\eqref{eq:disk_trap}. (a) Density
        of condensate 1, $n_1=|\psi_1|^2$. (b) Density of condensate 2, $n_2=|\psi_2|^2$. (c) Relative phase $\arg(\psi_1\psi_2^*)$.
        (d) Total phase $\arg(\psi_1\psi_2)$. The black dotted circles in panels (c) and (d) denote the trap boundary at the disk diameter of $2 L=70\xi$. 
        Other parameters are $\nu=2\times 10^{-3}\mu$, $g_{12}=0$.
        }
        \label{fig:2_lengthvm}
    \end{figure}

    Figure \ref{fig:2_lengthvm} shows the density and phase features of a vortex molecule from a numerical solution of the Gross-Pitaevskii equation (GPE). The component vortices are clearly distinguished by their low-density cores (dark dots) in panels (a) and (b). They also give rise to phase singularities (all colors of the rainbow meeting in a single point) in the phase plots in panels (c) and (d). 
    The domain wall is clearly visible as a region of large phase gradients in panel (c), showing the relative phase between the two condensates. The fact that the relative phase nearly vanishes outside of the localised domain wall indicates that both condensates have the same phase structure. This observation is inconsistent with the assumptions of the extended point vortex model of Refs.~\cite{Tylutki2016,Choudhury2022}, where the point vortices are located at different positions in the component condensates, namely at either end of the vortex molecule, which leads to a global misalignment of the component phase fields.
    In addition to the numerical observations, it also makes sense to assume that the phases of the component condensates align outside of the immediate vicinity of the domain wall, as this will minimise the local energy density \cite{Opanchuk2013}.

    Modelling instead the component vortices by a vorticity distribution that is equal in each component condensate and distributed along the domain wall, the phase of the two condensates is identical everywhere outside the domain wall. For the purpose of the distributed vorticity model we will assume a uniform vorticity distribution along the domain wall, which is modelled as a (narrow) straight line extending over the size of the molecule. The uniform distribution is the simplest assumption that can be made, and is furthermore consistent with a constant total phase of the component condensates, as it is observed in the GPE simulation, see Fig.~\ref{fig:2_lengthvm}(d). In essence, the assumption of a distributed vorticity along the molecular axis represents the action of the domain wall on the relative phase, but shrunken to a line of zero width.

    The assumed distributed vorticity is relevant for the vortex molecule dynamics by generating a continuous distribution of image vortices from the boundaries of the trap. In the following we consider a flat-bottom trap, which is modeled as a container with hard wall boundaries. 
    The concept of a distributed vorticity is thus used in a very different context than in Ref.~\cite{Groszek2018}, where distributed image vorticity was found useful in modelling  the vortex motion in a single condensate while dealing with a Thomas-Fermi parabolic density profile (or soft boundaries).

    Additional
    Magnus-force contributions to the dynamics originating from the domain-wall surface tension and core-interaction are obtained from the numerically generated vortex-molecule interaction energy as in our previous work~\cite{Choudhury2022}. For the current work we use a more accurate representation of the numerical data compared to the parameterization used in Ref.~\cite{Choudhury2022}, combining interpolation and extrapolation, which we found necessary in order to obtain quantitative agreement with fully numerical simulations of the vortex molecule dynamics. 
    The new parametrization is now consistent with the regimes of weak and strong Rabi coupling examined analytically in Ref.~\cite{Calderaro2017a}.

    We derive and solve the equations of motion for the distributed vorticity model for the case of a single vortex molecule in a disc shaped domain with hard wall boundary conditions. This situation could be achieved in BEC experiments with a flat-bottom trap. The model solutions are compared with fully numerical solutions of the GPE.
    We also compare with the simpler method of images for point vortices and a simplified description of the surface tension that is linear in the domain wall size and find that the refined model gives the best agreement with the GPE data.

    The paper is structured as follows. Section \ref{sec:meanfield} introduces the coupled GPE governing the system of coherently coupled
    condensates. 
    Section \ref{sec:pv} introduces the point vortex formulation and reviews the extended point vortex model before defining the distributed vorticity model.    
    The general formulations are applied to the dynamics of a vortex molecule in a flat-bottom trap in 
    Sec.~\ref{sec:pvframe} before concluding in Sec.~\ref{sec:conclusions}.
    Appendix \ref{appendixA} provides relevant details for the parameterization of the interaction energy and Appendix.~\ref{app:dist}
    discusses the calculation of the general integrals of the charge distribution model.

\section{Mean-Field Formulation} 
    \label{sec:meanfield}
    We characterize two coherently coupled atomic Bose-Einstein condensates with complex order parameters, $\psi_1(\mathbf{r},t)$ and $\psi_2(\mathbf{r},t)$ in the mean-field 
    description of  coupled GPEs
    \begin{subequations} \label{coupledGPE}
        \begin{align}
        i\hbar\frac{d\psi_1}{dt}&=\left(\hat{h} -\mu +g_1|\psi_{1}|^2 +g_{12}|\psi_{2}|^2 \right)\psi_1 -\nu \psi_{2}, \\
        i\hbar\frac{d\psi_2}{dt}&=\left(\hat{h} -\mu +g_2|\psi_{2}|^2 +g_{12}|\psi_{1}|^2\right)\psi_2 -\nu \psi_{1},
        \end{align}
    \end{subequations}
    where  $\hat{h}=-\frac{\hbar^2}{2m}\nabla^2+V_{\mathrm{ext}}$ is the single-particle Hamiltonian for bosons of mass $m$. The chemical potential $\mu$ controls the number of particles in numerical simulations. The spatially homogeneous coherent (Rabi) coupling $\nu$ can be realized by a two photon
    microwave field or a driving 
    radio frequency. 
    The atoms are assumed to be strongly confined along the $z$ spatial dimension to realize a quasi two dimensional quantum gas with a positional coordinate denoted as  $\mathbf{r} = (x, y)^\mathrm{t}$.
    Additional box-like confinement in the two-dimensional plane is provided by the external potential $V_{\mathrm{ext}}(\mathbf{r})$, which we take to vanish inside the superfluid domain and rising sharply at the domain boundaries.
    This will create a uniform quasi-two-dimensional Bose-Einstein condensate with rigid (hard-wall) boundaries as realized, e.g., in Ref.~\cite{Chomaz2015}.

    Our zero temperature theory is applicable to the low temperature and high particle-number-density regime of experiments \cite{Chomaz2015}.
    The free energy associated with the GPE~\eqref{coupledGPE} is given by the integral
    \begin{align}
          W= &\int \bigg[   \sum_{i=1}^2 \left(\psi_i^*\hat{h}\psi_i  + \frac{g_i}{2}|\psi_i|^4 -\mu |\psi_i|^2 \right) \nonumber \\ 
          & +g_{12}|\psi_1|^2|\psi_2|^2 -\nu (\psi_1^*\psi_2+\psi_1\psi_2^*)  \bigg]  \mathrm{d}^2 r. 
    \end{align} 

    In order to reduce the number of parameters, we choose  equal intra-component interactions $g_1=g_2\equiv g$.
    The homogeneous and time-independent ground-state solution of Eq.~\eqref{coupledGPE} for $V_{\mathrm{ext}}=0$ is given by equal and constant density $n_{1/2} \equiv |\psi_{1/2}|^2 = (\mu+\nu)/(g+g_{12})\equiv n_0$ in the component condensates. 
    We are interested in the miscible regime where 
    $g  + {|\nu|}/n_0 > g_{12}$ \cite{Abad2013}. The homogeneous solution $n_0$ serves as the background bulk density for localized vortex or non-linear-wave solutions. The linear coupling $\nu>0$ ensures that component condensates phases align and thus $\psi_1=\psi_2$ in equilibrium \cite{Brand2010}, but a global phase factor remains undetermined due to a global  $U(1)$ symmetry of the coupled BECs. The healing length 
    \begin{align}
        \xi=\frac{\hbar}{\sqrt{m(g+g_{12})n_0}}=\frac{\hbar}{\sqrt{m(\mu+\nu)}}
    \end{align}
    provides the length scale on which a homogeneous solution is recovered away from forced local 
    inhomogeneities due to vortex cores, or boundary conditions \cite{Pethick2008}.

    A point vortex model, nominally applicable to an incompressible fluid, requires that the healing length is smaller than other relevant length scales like the separation of vortices, or the distance of vortices from the boundaries \cite{Saffman1995,Skipp2022}. Reference \cite{Toikka2016} showed how to relax the incompressibility condition of the point vortex model and obtain correction terms for vortex dynamics as a series expansion in $\xi^2/D^2$, where $D$ is a length scale of the superfluid domain.
    Vortex molecules introduce two additional length scales on top of the healing length.
    The molecular size $d$ is the separation between the vortex singularities in the component vortices and also determines the length of the domain wall in the relative phase in situations where the domain wall extends along a straight line connecting the component vortices. The third length scale 
    \begin{align}
        \xi_\mathrm{J} = \frac{\hbar}{\sqrt{4m\nu}},
    \end{align}
    is called the Josephson vortex length scale and determines the width of the domain wall connecting the two component vortices. Exact solutions for a stationary and moving Josephson vortex were characterized in Refs.~\cite{Kaurov2005} and \cite{Shamailov2018}, respectively.

    In the numerical example shown in Fig.~\ref{fig:2_lengthvm} the molecular size $d=20\xi$ is slightly larger than that Josephson length scale $\xi_\mathrm{J}\approx 11.2\xi$. The domain wall of the relative phase is clearly seen as a feature with large phase gradients in the relative phase in panel (c). We can understand the domain wall to be centered around the line of constant relative phase $\arg(\psi_1\psi_2^*) = \pm \pi$ and extending over a width of $\xi_\mathrm{J}$.
    In the distributed vorticity model proposed in this work the domain wall is reduced to a straight line of zero width along which the vorticity of the vortex molecule is distributed. The model thus assumes that both the healing length $\xi$ and the Josephson length $\xi_\mathrm{J}$ are small compared to any other length scale, including the domain size $D$ and the molecular size $d$.

    The numerical vortex-molecule solution of the coupled GPE \eqref{coupledGPE} show in Fig.~\ref{fig:2_lengthvm} models 
    a disc-shaped two-component BEC.
    The disc-shaped radially-symmetric external potential of radius $L$ 
    is described by 
    \begin{align} \label{eq:disk_trap}
        V_{\mathrm{ext}}(\mathbf{r})=(\mu+\nu)\left(1+\tanh\frac{|\mathbf{r}|-L}{\xi}\right).
    \end{align}  
    The solution shown in Fig.~\ref{fig:2_lengthvm} was obtained by first imprinting a single vortex in each condensate component at the desired locations $\mathbf{R}_1$, $\mathbf{R}_2$.
    A low energy solution is then obtained by evolving  Eq.~\eqref{coupledGPE} in imaginary time, i.e.~replacing $t\to -i\tau$, which corresponds
    to minimizing the energy functional $W$ by gradient flow.
    Imaginary time evolution quickly removes most density  and non-topological phase excitations. On a slower timescale, the location of vortex phase singularities move towards lower energy configurations, which eventually moves them outside the trap.
    To avoid this, Gaussian pinning potentials are used to pin the vortices in 
    a particular configuration for each vortex while only having minimal effect on the phase and density structure. 
    While the obtained solutions are stationary only in the presence of the pinning potential, they also serve as suitable initial conditions for studying vortex dynamics under real-time evolution of the coupled GPE \eqref{coupledGPE} after the pinning potentials are removed. Under the assumptions of the point vortex model, the vortex dynamics only depends on the instantaneous position of the vortex singularities by evolving through minimal energy configurations. 
    This neglects, in particular, the effects of sound emission or reabsorption.
    We test the predictions of the point vortex model by comparing with fully time-dependent GPE dynamics. Vortex positions are tracked by accurately locating the phase singularities using the software library \texttt{VortexDistributions.jl} \cite{Bradley2022}.

    
\section{Point vortex formulation} \label{sec:pv}

The idea of the point vortex model is that the dynamics of vortices is fully determined by the positions of all vortices in the system together with the boundary conditions. The model strictly applies to ideal inviscid and incompressible fluids \cite{Saffman1995,Newton2001}, and can be applied to the GPE in situations where the healing length $\xi$ can be considered a small parameter \cite{Toikka2016,Skipp2022}. For an ideal fluid in two dimensions one can define a scalar stream function whose contours are co-linear with the local fluid velocity $\mathbf{u}(\mathbf{r})$. 
The velocity of a point vortex at position $\mathbf{R}$
can be found from
the stream function
after the singular contribution of the vortex itself has been removed \cite{Newton2001}.
While the stream function cannot be used for multi-component coupled BECs,
an alternative approach based on energy conservation is still applicable.

The idea is to find the point vortex trajectories as the contours of a conserved energy function, which only depends on the vortex coordinates after the boundary conditions are defined. This leads to a Hamiltonian formulation where the vortex $x$ and $y$ coordinates play the role of canonically conjugate variables. In this formulation  additive contributions to the total energy  provide additive contributions to the vortex velocity.

\subsection{Extended point vortex model for the vortex molecule} \label{sec:ext_p_v_m}
In Ref.~\cite{Choudhury2022} we presented an extended point vortex model where the total energy of a vortex molecule is given by 
\begin{align} \label{eq:Etot}
    E_\mathrm{vm}(\mathbf{R}_1, \mathbf{R}_2) = E_\mathrm{bound}(\mathbf{R}_1, \mathbf{R}_2)  +
    V(|\mathbf{R}_1- \mathbf{R}_2|),
\end{align}
where $\mathbf{R}_j = (X_j, Y_j)^\mathrm{t}$ is the coordinate vector of the $j$th component vortex,  $E_\mathrm{bound}(\mathbf{R}_1, \mathbf{R}_2)$ is an energy contribution from the boundary-induced image vortices, and $V(d)$ is an internal energy of the vortex molecule that only depends on the separation of the component vortices, or the molecular size, $d = |\mathbf{R}_1- \mathbf{R}_2|$. The trajectories of the component vortices are then obtained from the equation of motion
\begin{align} \label{eq:eom}
    \dot{\mathbf{R}}_j &= \nabla^\perp_j 
    \frac{E_\mathrm{vm}(\mathbf{R}_1, \mathbf{R}_2)}{2\pi n_0 \hbar \kappa}, \\
\end{align}
where $\kappa = \pm 1$ is the integer vortex charge and 
\begin{align}
    \nabla^\perp_j = \begin{pmatrix} \frac{\partial}{\partial Y_j} \\ -\frac{\partial}{\partial X_j} \end{pmatrix} 
\end{align}
is the projection of the curl onto the $x$--$y$ plane. 
The equation of motion \eqref{eq:eom} has the structure of Hamilton's equations of motion common to the Hamiltonian formulation of point vortex dynamics \cite{Newton2001}. As a consequence of the two energy contributions of Eq.~\eqref{eq:Etot} the equation of motion has two contributions to the point vortex velocity:
\begin{align}
    \label{eq:epvm}
    \dot{\mathbf{R}}_j &= \mathbf{V}^\mathrm{bound}_{j} + \mathbf{V}^\mathrm{int}_{j}.
\end{align}

Reference~\cite{Choudhury2022} made specific assumptions for the two terms in Eq.~\eqref{eq:Etot}: The boundary term was provided by the method of images for each component vortex separately
\begin{align}
    E_\mathrm{bound}(\mathbf{R}_1, \mathbf{R}_2) = E_\mathrm{sv}(\mathbf{R}_1) + E_\mathrm{sv}(\mathbf{R}_2),
\end{align}
where $E_\mathrm{sv}(\mathbf{R})$ is the energy contribution of a single vortex in the superfluid within the given boundaries, i.e.~incorporating the contributions from image vortices.
As a consequence, 
the point vortex velocity contributions in the equation of motion become
\begin{align}
    \label{eq:e_sv}
    \mathbf{V}^\mathrm{bound}_{j} &= \mathbf{V}^\mathrm{sv}_{j} \equiv ({2\pi n_0 \hbar \kappa})^{-1} \nabla^\perp E^\mathrm{sv}(\mathbf{R})|_{\mathbf{R}=\mathbf{R}_j}\\
    \label{eq:e_int}
    \mathbf{V}^\mathrm{int}_j &= ({2\pi n_0 \hbar \kappa})^{-1}\frac{\mathrm{d}V(d)}{\mathrm{d}d} \nabla^\perp_j |\mathbf{R}_1 - \mathbf{R}_2|.
\end{align}
The interaction energy $V(d)$ was parameterized from a numerical calculation of the vortex-molecule energy in the absence of boundaries.
This model provided an adequate description of the dynamics of the vortex molecule in a channel with parallel side walls and was able to reproduce all qualitative phase space structures in Ref.~\cite{Choudhury2022}.

We note, however, that this model is too simplistic for a fully quantitative description of vortex-molecule dynamics.
In particular, the velocity contribution of the boundary term $\mathbf{V}^\mathrm{sv}_{j}$ originates from the image vortices of the component vortex at position $\mathbf{R}_j$ only. We know however that the phase structure of
the component condensates is not independent of each other but rather 
strongly influenced by the Rabi coupling. 
A domain wall of the relative phase extends between the  component vortices roughly along the molcular axis.
At distances larger than $\xi_\mathrm{J}$ away from the  
domain wall, the relative phase drops to zero and the phases of each component BECs align, see Fig.~\ref{fig:2_lengthvm}. 

This observation motivates us to modify the extended point vortex model by considering an image vortex distribution extended along the image of the domain wall of the relative phase.
For this work we retain the formulation of the extended point vortex model of Eqs.~\eqref{eq:Etot} and \eqref{eq:eom}, but improve the approximate representation of the two energy terms. For the interaction energy $V(d)$ we use a more accurate numerical representation based on the same numerical calculation as detailed in Appendix \ref{appendixA}. An improved representation of the boundary contributions to the vortex molecule equation of motion is the subject of the distributed vorticity model.

\subsection{Distributed vorticity model}

Due to the effect of  the boundaries each vortex obtains a velocity component that can be thought of as the linear superposition of velocity fields induced by all image vortices at the position of that vortex. Let 
\begin{align}
    \kappa \mathbf{u}_\mathrm{im}(\mathbf{r}; {\mathbf{R}})
\end{align}
denote the velocity field induced by the images of a vortex at position $\mathbf{R}$ with charge $\kappa$.
The boundary-induced velocity contribution in the extended point vortex model of Sec.~\ref{sec:ext_p_v_m} of vortex $j$ then becomes 
\begin{align} \label{eq:v_sv}
    \mathbf{V}_j^\mathrm{sv} = \kappa \mathbf{u}_\mathrm{im}(\mathbf{R}_j; {\mathbf{R}_j}).
\end{align} 

\begin{figure}
    \centering
    \includegraphics[width=0.8\linewidth]{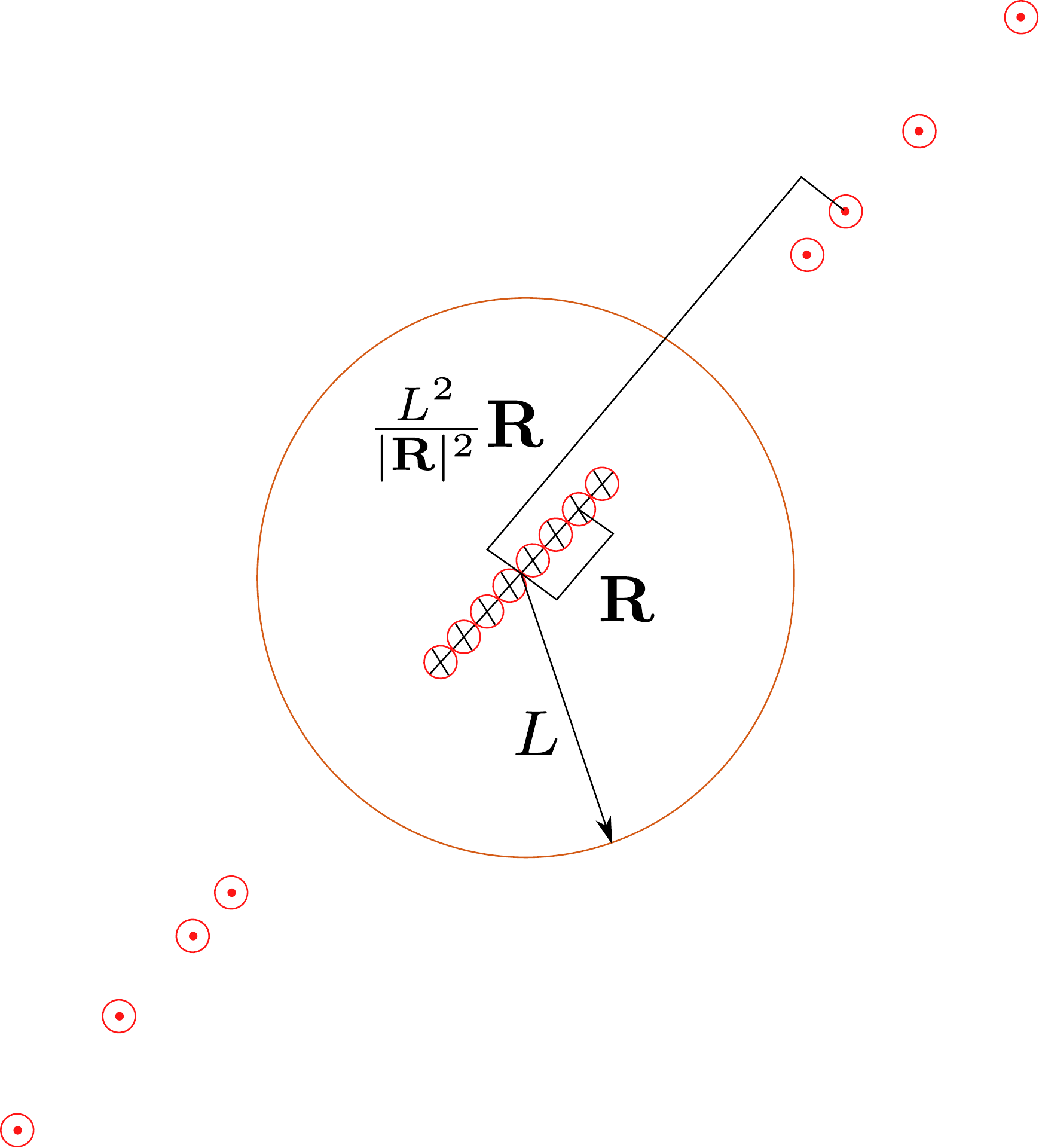}
    \caption{Concept diagram of the vorticity distribution model for a centered vortex molecule in a disc trap. The image vortices are formed at locations $\tilde{\mathbf{R}}=\frac{L^2}{|\mathbf{R}|^2} \mathbf{R}$ where
    $\mathbf{R}$ is the position of the original charge. The distance of the image vortices from the center of the condensate is inversely
    proportional to the distance of its original vortex from the center. Hence, even though the charge distribution is evenly spaced the image charges are not and reach out to infinity as $|\mathbf{R}|\to 0$.}
    \label{concept_charge}
\end{figure}

For the distributed vorticity model we take the source of image vorticity to be distributed along the domain wall of the relative phase extending between the two component vortices of the vortex molecule, i.e.~along the molecular axis.
Thus the velocity component of vortex $j$ originating from the image vorticity becomes a superposition of velocity contributions 
\begin{align} \label{eq:v_dv}
    \mathbf{V}_j^\mathrm{dv} = \kappa \int_0^1 \mathbf{u}_\mathrm{im}(\mathbf{R}_j; {(1-t)\mathbf{R}_1 + t \mathbf{R}_2})  \mathrm{d}t.
\end{align}
Figure \ref{concept_charge} is a concept diagram that shows how the distributed vorticity of the vortex molecule can be thought of as being composed of individual vortices of fractional charge, giving rise to a distribution of image vortices in turn. 
The placement of image vortices is chosen such that the no-flow boundary condition for flow perpendicular to the disk boundary is satisfied.
In the following we will apply these ideas to the disc geometry.

\section{Vortex molecule dynamics in a flat-bottom disc trap}
\label{sec:pvframe}
\subsection{Velocity of a simple vortex in a disc}
The simple vortex solution in the two-component condensate can be understood as a special case of a vortex molecule where the two vortices are at the same location. In this case the phases of the two component condensates can align perfectly and thus no domain wall of the relative phase is present. The velocity of the simple vortex in a disc is thus completely determined by the contribution from the boundaries. Moreover, the predictions from the extended point vortex model and from the distributed vorticity model trivially agree and become equivalent to the point vortex model for a single-component superfluid. 

In the point vortex model we ignore the compressibility of the superfluid, formally taking $\xi\to 0$, and assume a constant condensate density. 
The phase of the GP order parameter of a single vortex at position $\mathbf{R}=(X,Y)^\mathrm{t}$ is given by $\arg[\psi(\mathbf{r})]=\kappa\arctan\frac{y-Y}{x-X}$ in the absence of boundaries (up to a constant), and the correponding velocity field is
\begin{align}
    \mathbf{u}(\mathbf{r})=\frac{\hbar}{m} \nabla \arg({\psi})= \frac{\hbar \kappa}{m}\frac{\hat{z}\times (\mathbf{r}-\mathbf{R})}{|\mathbf{r}-\mathbf{R}|^2}.
\end{align}
The influence of the box-like trapping potential is  to create a no-flow boundary condition, i.e.~a condition that prohibits a perpendicular component of the superfluid velocity distribution at the boundary.
For a circular disc centered at the origin with radius $L$ this boundary condition is met by adding an image vortex with charge $-\kappa$ at the position \cite{Newton2001}
\begin{align}
    \tilde{\mathbf{R}}=\frac{L^2\mathbf{R}}{|\mathbf{R}|^2} .
\end{align}
The velocity field induced by the image of a vortex with charge $\kappa$ at position $\mathbf{R}$ is thus 
\begin{align} \label{eq:disc_image}
    \kappa \mathbf{u}_\mathrm{im}(\mathbf{r}; {\mathbf{R}}) = -\frac{\hbar \kappa}{m}\frac{\hat{z}\times (\mathbf{r}-\tilde{\mathbf{R}})}{|\mathbf{r}-\tilde{\mathbf{R}}|^2}.
\end{align}

The boundary contributions to the velocities of the component vortices in a vortex molecule can now be obtained by substituting Eq.~\eqref{eq:disc_image} into Eqs.~\eqref{eq:v_sv} and \eqref{eq:v_dv} for the extended point-vortex model and the distributed vorticity models, respectively. The full equation of motion for the vortex molecule is then given by Eq.~\eqref{eq:epvm} in combination with Eq.~\eqref{eq:e_int}.

\subsection{Precession of a centered vortex molecule}

In order to quantitatively compare between the different models, we now focus on the situation where the vortex molecule is located symmetrically in the center of the disc with $\mathbf{R}_1 = - \mathbf{R}_2$ and $|\mathbf{R}_j| = d/2$. In this case the symmetry is preserved during the vortex motion. The point vortex velocity is perpendicular to the molecular axis and the vortex molecule rotates with a constant precession frequency around the axis of the disc trap
\begin{align}
    \nonumber
    \Omega_\mathrm{vm} \hat{\mathbf{z}} &= \frac{\mathbf{R}_j \times \mathbf{V}_j}{|\mathbf{R}_j|^2}, \\ \label{eq:Omega_vm}
    &= (\Omega_\mathrm{bound} + \Omega_\mathrm{int}) \hat{\mathbf{z}} ,
\end{align}
which breaks up into components originating from the boundary and interactions as per Eq.~\eqref{eq:epvm}.
The interaction contribution to the precession frequency becomes, from Eq.~\eqref{eq:e_int}
\begin{align} \label{eq:Omega_int}
    \Omega_\mathrm{int} = \frac{\kappa}{\pi n_0 \hbar d} \frac{\mathrm{d}V(d)}{\mathrm{d}d}.
\end{align}

In a situation where the domain  wall energy is purely linear in $d$, as derived for $d \gg \xi_\mathrm{J}$ in Ref.~\cite{Calderaro2017a}, the gradient term is constant and the interaction contribution to the precession frequency becomes inversely proportional to the molecular length scale $d$. This term is divergent for small $d$.

\begin{figure}
    \centering
    \includegraphics[width=1\linewidth]{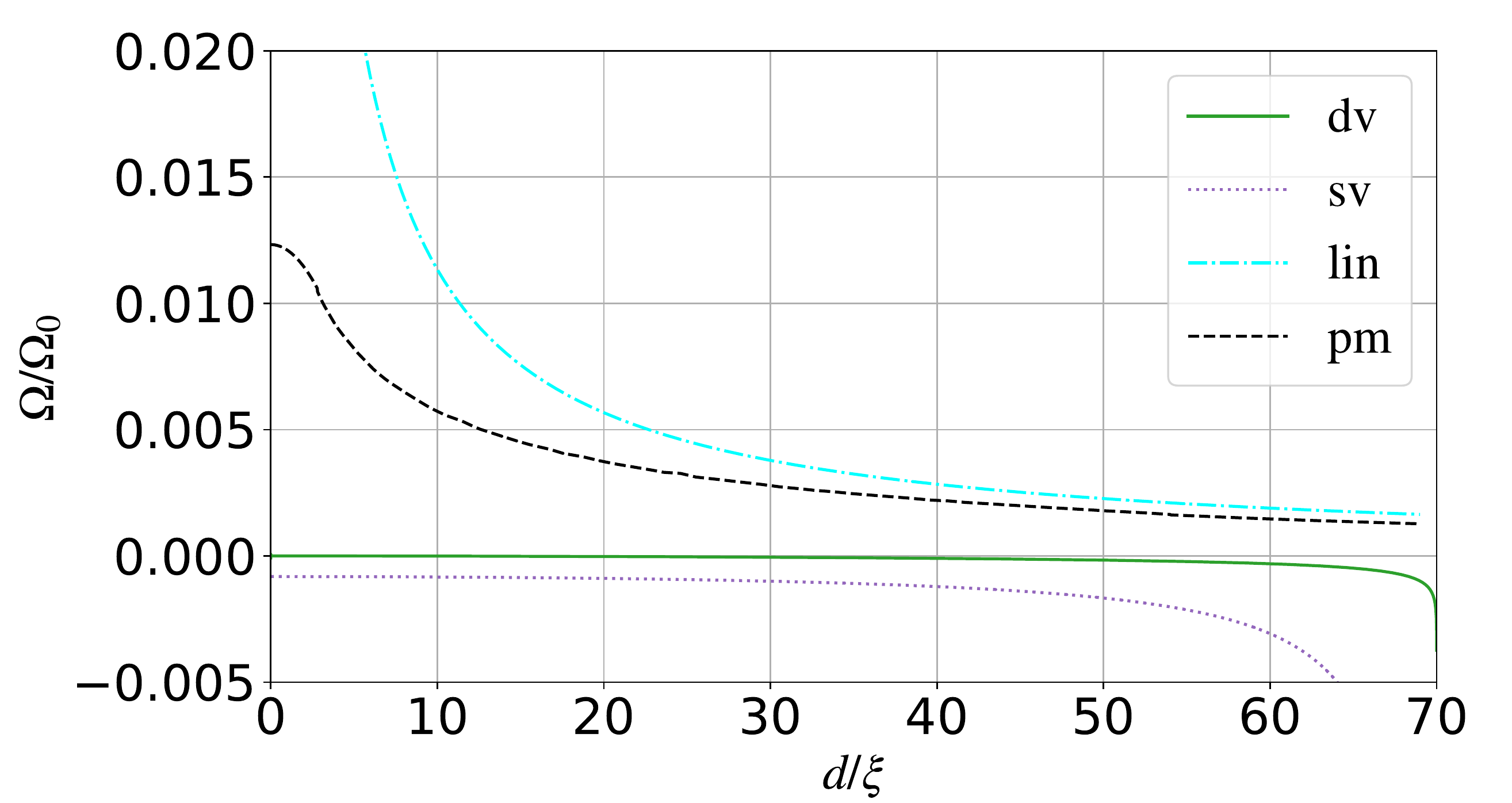}
    \caption{
    Components of the vortex molecule precession frequency according to various models as a function of the molecular size $d$. The negative-valued boundary contributions $\Omega_{\mathrm{bound}}$ are labeled ``$\mathrm{dv}$'' for the distributed velocity contribution $\Omega_{\mathrm{dv}}$ of Eq.~\eqref{eq:cd_frequency} and ``$\mathrm{sv}$'' for the single vortex contribution $\Omega_{\mathrm{sv}}$ of Eq.~\eqref{eq:sv_frequency}.
    The interaction components $\Omega_{\mathrm{int}}$ follow Eq.~\eqref{eq:Omega_int} and bring positive contributions. The one labeled ``$\mathrm{lin}$'' follows from a purely linear interaction potential with the surface tension of the idealized Josephson vortex. The contribution from the parameterized numerically obtained interaction energy is labelled ``$\mathrm{pm}$''.
    The frequency is expressed in units of $\Omega_0=(\mu +\nu)/\hbar$. Parameters
    are same as Fig.~\ref{fig:2_lengthvm} }
    \label{omega_components}
\end{figure}

For the boundary contribution $\Omega_\mathrm{bound}$ we consider the single vortex contribution from the extended point vortex model and the distributed vorticity model. The precession frequency from the extended point vortex model becomes [from Eqs.~\eqref{eq:v_sv} and \eqref{eq:disc_image}]
\begin{align} \label{eq:sv_frequency}
    \Omega_\mathrm{sv}(d) = -\frac{\hbar \kappa}{m} \frac{4}{4L^2 - d^2}.
\end{align}
This is the result  of Ref.~\cite{Choudhury2022}.
Obtaining the precession frequency of the distributed vorticity model requires evaluating the integral in Eq.~\eqref{eq:v_dv}. A closed form solution can be found and leads to 
\begin{align}
    \label{eq:cd_frequency}
    \Omega_{\mathrm{dv}}(d)=\frac{\hbar \kappa }{m}\frac{4}{d^4}\left[d^2-2L^2\ln\left(\frac{4L^2+d^2}{4L^2-d^2}\right)\right] .
\end{align} 
Expanding in powers of $d$ in the vicinity of $d=0$ we obtain
\begin{align}
    \label{eq:lin_intfreq}
    \Omega_{\mathrm{dv}}(d)= -\frac{\hbar \kappa }{m}\frac{d^2}{12L^4}+\mathrm{O}(d^6),
\end{align}
where the leading term is of second order in $d$.
Thus, the boundary contribution to the precession frequency from the distributed vorticity model vanishes at small molecular size. This is in contrast to the single-vortex contribution from the extended point vortex model of Eq.~\eqref{eq:sv_frequency}, which has the finite limit $\Omega_\mathrm{sv}(0) = -{\hbar \kappa }/{mL}$.
A comparison of the different contributions to the precession frequency is shown in Fig.~\ref{omega_components}.

Figure \ref{omega_components} also shows two different curves for $\Omega_\mathrm{int}$ according to different models for the interaction energy of the vortex molecule. The simplest choice with a linear $d$ dependence is 
\begin{align} \label{eq:V_lin}
    V_\mathrm{lin}(d) = d \,\sigma,
\end{align}
where 
\begin{align} \label{eq:sigmaJ}
    \sigma =  \frac{8\hbar \sqrt{\nu}}{3\sqrt{m}} \frac{3\mu-\nu}{g+g_{12}},
\end{align}
is the energy (line density) of a Josephson vortex  \cite{Kaurov2005}, the exact solution for a non-moving domain wall of the relative phase. Reference \cite{Tylutki2016} used this model with an approximate domain wall energy density from Ref.~\cite{Son2002} valid for small $\nu$, which was also derived for $d\gg \xi_\mathrm{J}$ in Ref.~\cite{Calderaro2017a}.

As an alternative we have parameterized the numerically computed vortex molecule interaction energy $V_\mathrm{pm}(d)$. The numerical calculation of the interaction energy was first reported in Ref.~\cite{Choudhury2022}. For the current work we have re-parameterized the numerical data in order to obtain increased accuracy as detailed in Appendix~\ref{appendixA}. 
As an alternative we have parameterized the numerically computed vortex molecule interaction energy $V_\mathrm{pm}(d)$. The numerical calculation of the interaction energy was first reported in Ref.~\cite{Choudhury2022}. For the current work we have re-parameterized the numerical data in order to obtain increased accuracy as detailed in Appendix~\ref{appendixA}. 
We find that the parameterized numerical interaction energy as well as the derived frequency contribution $\Omega_\mathrm{pm}$ are significantly smaller than the linear model and deviate from it quite strongly for small and moderate values of $d$ while it asymptotically agrees at large $d$, as is expected.

\begin{figure}
    \centering
    \includegraphics[width=1\linewidth]{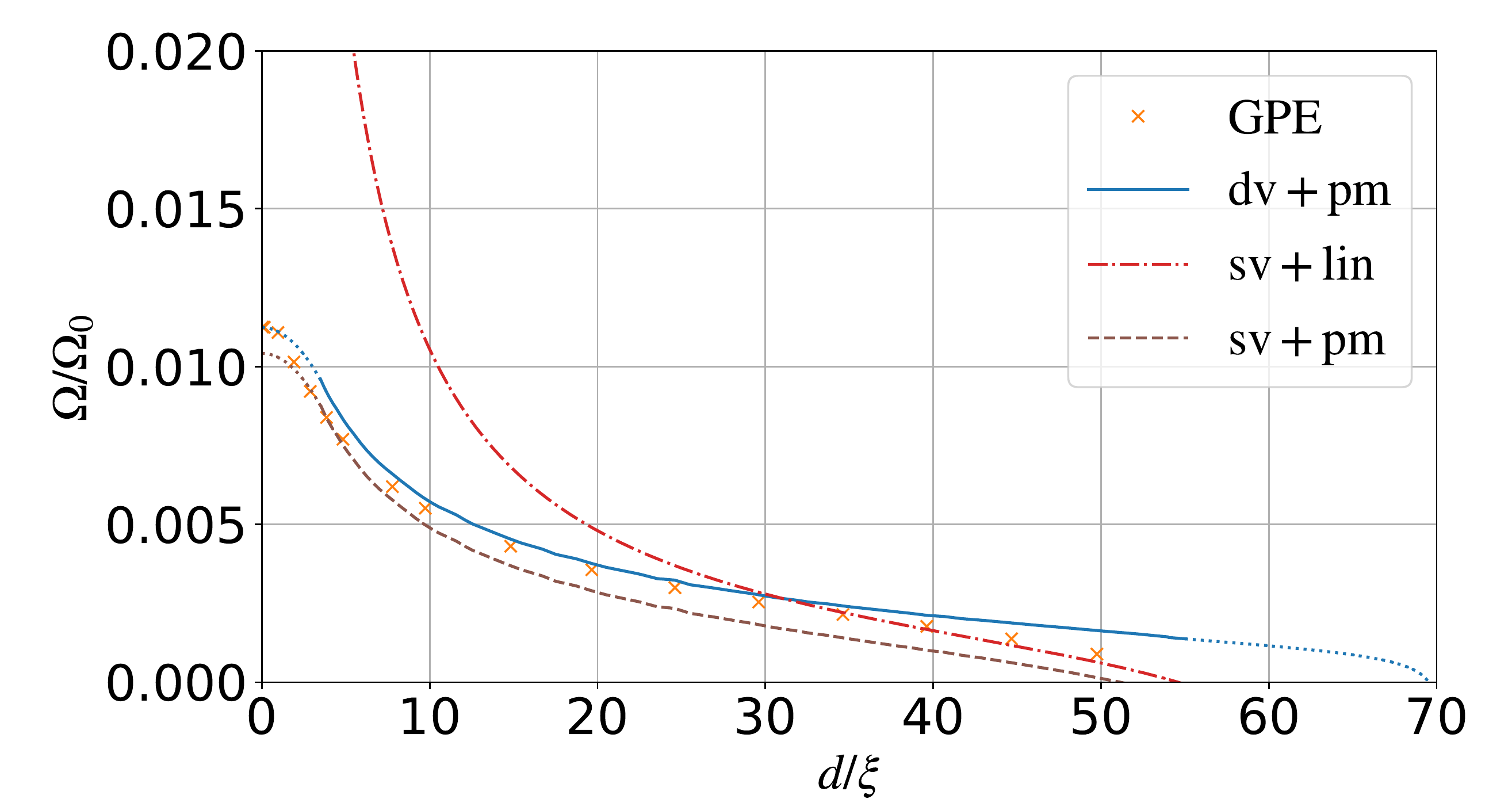}
    \caption{Precession frequency of a centered vortex molecule in a disc trap as a function of the molecular size $d$.
    The orange crosses represent real time evolution data from the GPE of Eq.~\eqref{coupledGPE}.  
    The curves are model predictions according to Eq.~\eqref{eq:Omega_vm} with different combinations 
    of the contributions for $\Omega_\mathrm{bound}$ and $\Omega_\mathrm{int}$  shown in Fig.~\ref{omega_components}. The blue solid line marked ``dv+pm''
    combines the distributed vorticity variant of the boundary contribution with the parametrized interaction energy and provides the best explanation of the numerical GPE data within the available models.
    The red dot-dashed line marked ``sv+lin'' represents single vortex boundary contribution from Eq.~\eqref{eq:sv_frequency} along with the linear interaction energy contribution from Eq.~\eqref{eq:V_lin}.
    The brown dashed line marked ``sv+pm'' represents single vortex boundary contribution along with parametrized interaction energy contribution.  
    The frequency is expressed in units of $\Omega_0=(\mu +\nu)/\hbar$. Parameters
    are same as Fig.~\ref{fig:2_lengthvm} }
    \label{omega_comparisonvsgpe}
\end{figure}

In Fig.~\ref{omega_comparisonvsgpe} we show numerical results for the vortex molecule precession frequency from GPE simulations in comparison with model predictions combining different components for the boundary and interaction contributions.
For small molecular size $d$ the values of the precession frequency are dominated by the interaction contributions. The linear interaction energy model leads here to a divergent contribution, which is unphysical. The parameterized interaction contribution however, captures the numerically observed finite precession frequency at small molecular sizes rather nicely. Up to intermediate molecular sizes compared to the disk radius of $L=35\xi$ the distributed vorticity and the single vortex contributions differ by an approximately constant shift with the gap widening for larger $d$. Over a wide range of molecular distances, the distributed vorticity model provides a better match with the GPE simulation data compared to the single vortex image model.

\begin{figure}
    \centering
    \includegraphics[width=1\linewidth]{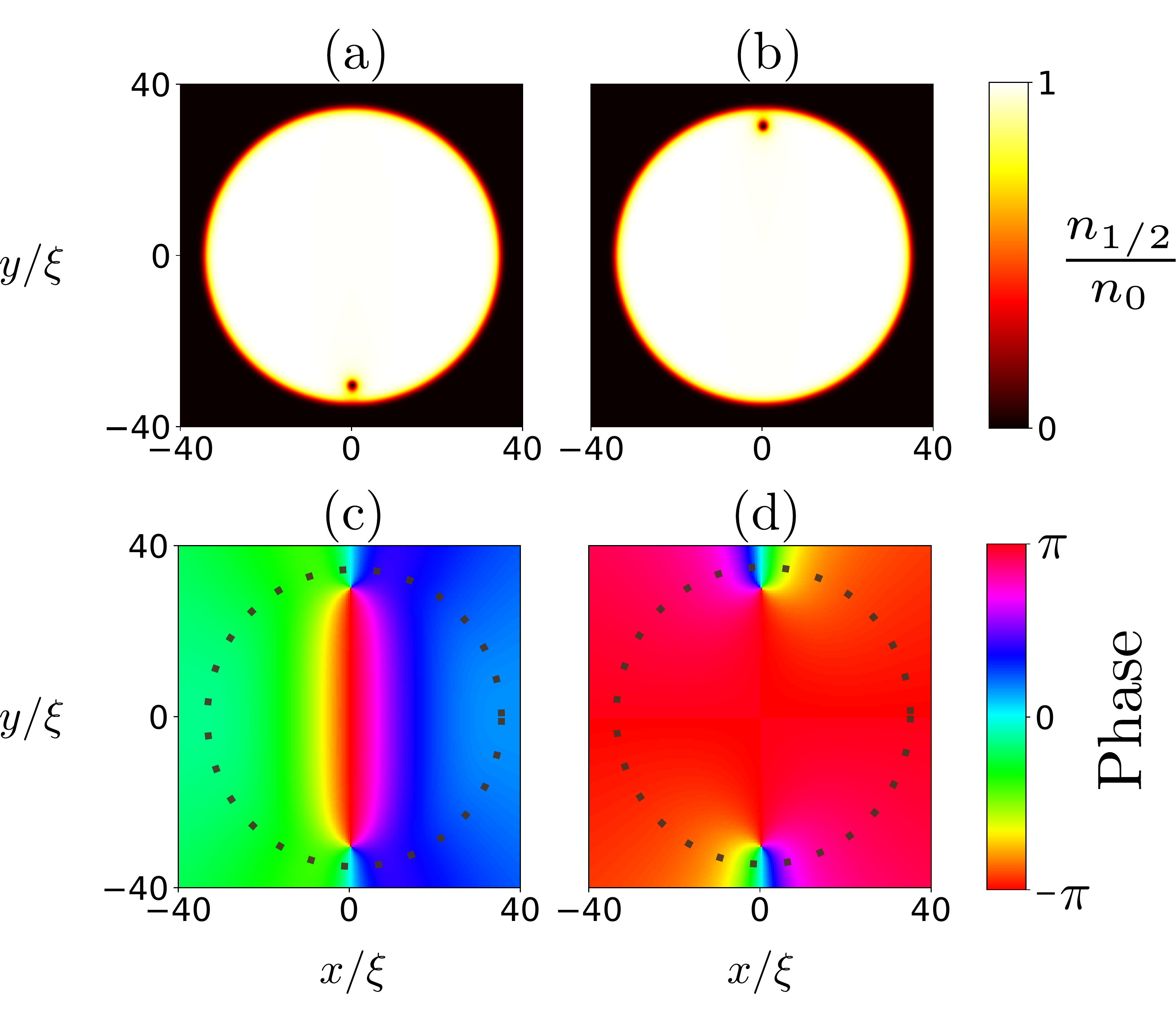}
    \caption{Numerical solution of the GPE with a centered vortex molecule with molecular length $d=60\xi$. The component vortices are separated less than $\xi_\mathrm{J}$ from the trap boundaries and the assumption of a localized domain wall breaks down. (a) Density
    of condensate 1 $n_1 = |\psi_1|^2$. (b) Density of condensate 2 $n_2 = |\psi_2|^2$. (c) Relative phase $\arg(\psi_1\psi_2^*)$.
    (d) Total phase $\arg(\psi_1\psi_2)$. Parameters as in Fig.~\ref{fig:2_lengthvm}.}
    \label{fig:60lenghtvm}
\end{figure} 

At large molecular size, where $d$ becomes comparable to the disk diameter $2 L$, the otherwise favored model ``dv+pm'' develops discrepancies from the GPE simulations seen in Fig.~\ref{omega_comparisonvsgpe}. In order to rationalize the failure of the model in this regime we visualize the numerical solution of the GPE for a large vortex molecule in Fig.~\ref{fig:60lenghtvm}. The component vortices are close to the boundaries of the disk trap in this case. While the relative phase in panel (c) clearly shows a strong feature reminiscent of a domain wall in the relative phase, it can also be seen that the relative phase does not return to values close to zero outside a localized region but rather differs from zero for most of the superfluid domain. Thus a crucial assumption of our model, i.e.\ the existence of a localized domain wall of the relative phase is violated for this case. The nonzero relative phase in the coupled BECs indicates that tunnel currents are present throughout the trap
due to the Josephson-like relation between current and phase in linearly coupled Bose-Einstein condensates \cite{Smerzi1997}. One of the consequences is that the continuity equation for each individual component, which underlies the point-vortex model, is now violated throughout the trap. The proximity of the component vortices to the trap boundary thus invalidated the assumptions of the distributed vorticity model and explains the discrepancies between the GPE data and model predictions for the precession frequencies seen in Fig.~\ref{omega_comparisonvsgpe} for $d \gtrsim 40\xi$.

\subsection{Off-centered vortex molecule}
     
\begin{figure}
    \centering
    \includegraphics[width=0.95\linewidth]{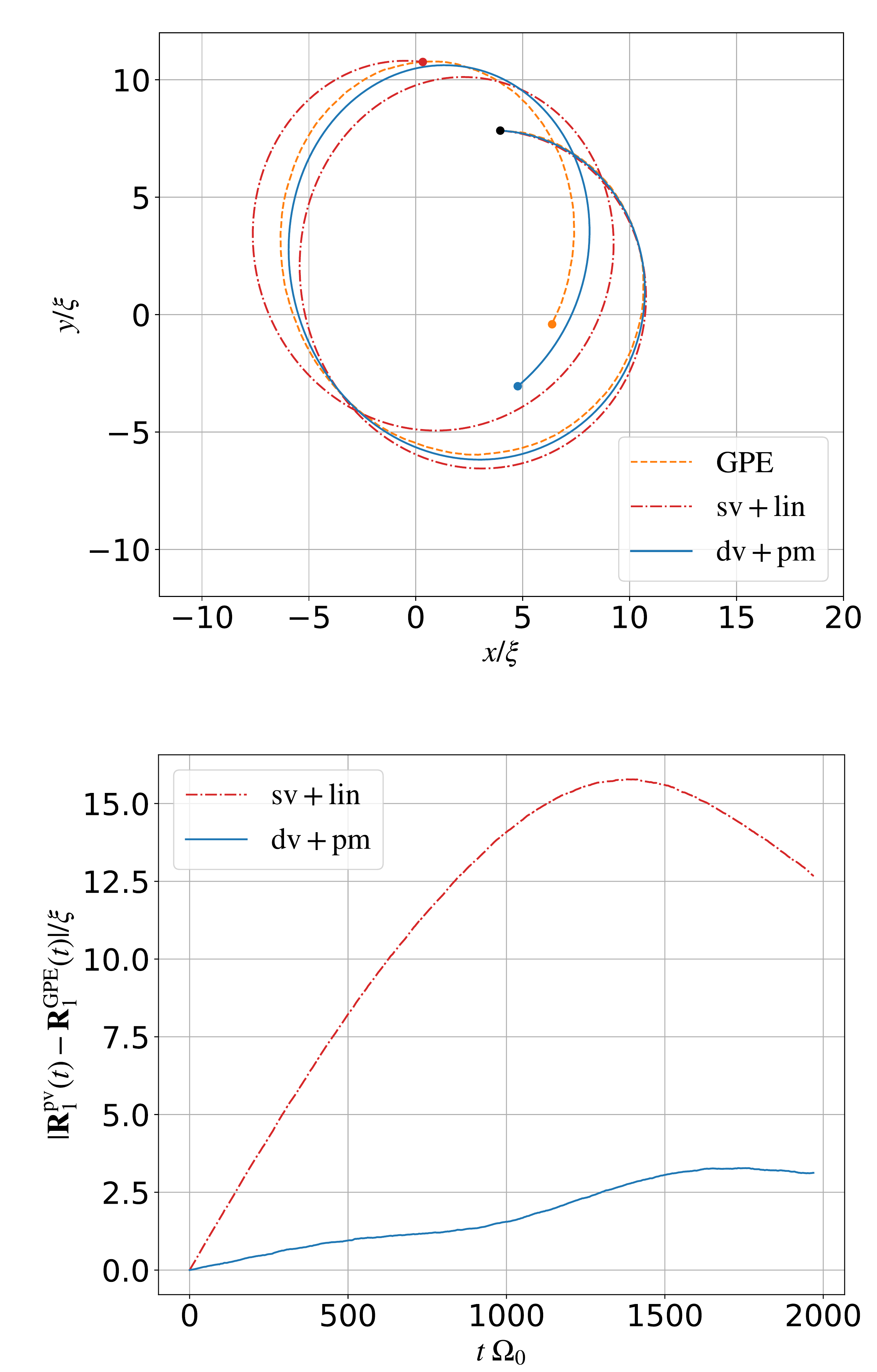}
    \caption{Trajectories predicted for an off-centered vortex molecule by different models, and their deviations.
    An off-centered vortex molecule moves in a disk-shaped trap ($L=30\xi$) with initial positions $\mathbf{R}_1(t=0) = (3.95\xi,7.83\xi)$ (black dot in panel (a)), and $\mathbf{R}_2(t=0) = (2.07\xi,-7.71\xi)$ (not shown).
    (a) Trajectories. The trajectory of vortex 1 is 
    represented by colored lines showing the predictions from different models as in Fig.~\ref{omega_comparisonvsgpe}. The dashed line ``GPE'' denotes the numerical solution of the full Eq.~\eqref{coupledGPE}. The dash-dotted line ``sv+lin'' shows the trajectory for the model with a single image vortex contribution of Eq.~\eqref{eq:sv_frequency} combined with a linear interaction contribution based on Eq.~\eqref{eq:V_lin}. The full line ``dv+pm'' is obtained from combining the distributed vorticity contribution of Eq.~\eqref{eq:sv_frequency} with the parameterized interaction contribution from Appendix~\ref{appendixA}. The colored dots mark the endpoints of the respective trajectories.
    (b) Deviations of the point vortex models from the GPE time evolution shown in panel (a). The instantaneous distance $|\mathbf{R}_1^\mathrm{pv}(t) - \mathbf{R}_1^{\mathrm{GPE}}(t)|$ between the location of vortex 1 according to the point vortex model and the GPE simulation is plotted as a function of time. Results are shown for the two point vortex models of panel (a) and labelled ``sv+lin'' and  ``dv+pm'', respectively. The unit of time is $1/\Omega_0=\hbar/(\mu +\nu)$.}      
    \label{off_centre_traj}
\end{figure} 
    
In addition to the centered vortex molecule, our models can also predict the trajectories of non-centered vortex molecules using the more general equation of motion \eqref{eq:eom} with details worked out in Appendix~\ref{app:dist}.
Figure \ref{off_centre_traj} shows the trajectory of one component vortex of a non-centered vortex molecule with $d=16\xi$. 
The fact that trajectories do not overlap for the different models as seen in Fig.~\ref{off_centre_traj}(a) indicates that the model predictions differ by more than just the precession frequency.
Comparing the GPE trajectory with the model solution we find that the model combining a single vortex image with the linear assumption for the interaction energy (marked ``sv+lin'') not only severely overestimates the angular rotation frequency but also produces deviations from the correct trajectory. 
Figure 6(b) further visualizes the deviations of the point-vortex model from the GPE trajectory by plotting the instantaneous distance of vortex 1 between the point-vortex model and the GPE trajectory as a function of time.
In comparison, the model based on distributed image vorticity combined with the parameterized interaction energy compares much more favorably to the GPE trajectories.

\section{Conclusion}
\label{sec:conclusions}

In conclusion, we have presented a distributed vorticity model for the dynamics of vortex molecules in a two-component Bose-Einstein condensate with linear coherent coupling. 
Our model extends previous work by considering a continuous distribution of image vorticity reflecting the effect of the domain wall on the vortex molecule phase structure. 
Specifically, the distributed vorticity model predicts a quadratic dependence for the image-induced contribution to the precession frequency on the length of the domain wall for small vortex molecules [Eq.~\eqref{eq:lin_intfreq}], while previous extended point vortex models predicted a constant angular frequency in the small molecule limit [Eq.~\eqref{eq:sv_frequency}].
A second major finding is that assumption of a linear interaction energy made in Ref.~\cite{Tylutki2016} leads to an unphysical divergence in the precession frequency that is inconsistent with the GPE data, while our model with the improved parametrization of the interaction energy avoids this unphysical divergence, and is consistent with the findings of Ref.~\cite{Calderaro2017a}. 
We tested our model predictions over a range of molecule sizes against numerical simulations in a two-dimensional circular disc and found support for the improved model.

The main benefit of the distributed vorticity model compared to full GPE simulations is that it provides conceptual insights into the dynamics of vortex molecules. An extension to the dynamics of multiple vortex molecules and their interactions should be possible and can be expected to work well as long as the vortex molecules are well separated. This can be assured in a low-density and low-energy regime due to the linear-in-length energy content of the domain wall. Such a model could be useful for the study of quantum turbulence with a large number of vortex molecules in coupled BECs. Solving the coupled ordinary differential equations for the distributed vorticity model can be done with less computational effort than solving the full coupled GPEs. Moreover the absence of domain wall reconnections and vortex anniliation in the distributed vorticity model may provide key insights into the importance of such features for macroscopic observables when comparing the results to GPE simulations where all of this is included.

Our findings contribute to the ongoing research on the dynamics of superfluid vortices, which still poses many open questions. The peculiar structure of vortices in multi-component BECs, the competition between Rabi coupling and nonlinear mean-field energy, and the analogies to axions and quark confinement make this a rich and active area of research.

In this work we have specifically considered the case of vanishing cross-component nonlinearity in the GPE model, i.e.~$g_{12}=0$. In the more general case where $g_{12} > 0$ the cores of the component vortices are partly filled by local density maxima of the other component \cite{Choudhury2022,Tylutki2016}. This leads to a more complicated dynamics that could be modelled, at the point vortex level, by including inertial effects as in Refs.~\cite{Richaud2020,Richaud2021}. 
Combining these ideas with the distributed vorticity model presented here could lead to a more accurate description of the dynamics of vortex molecules in a multi-component BEC with cross-component interactions.

\section{Acknowledgements}

S.C. acknowledges support by the Dodd-Walls Centre for Photonic and Quantum Technologies through a Dodd-Walls Centre PhD scholarship.

\appendix
\section{Interaction energy}
\label{appendixA}
The interaction energy is found as a function of molecular length $d$ by creating a vortex molecule 
in a large computational domain of $180\xi \times 180\xi$ 
and 
using imaginary time evolution of the GPE in Eq.~\eqref{coupledGPE}, i.e.~replacing the time variable $t$ by $-i\tau$, as described in Ref.~\cite{Choudhury2022}. 
The imaginary time evolution continually deforms the solution towards lower energy. While locally adjusting the correct density and gradients happens fairly quickly, on a slower time scale the position of the component vortices and thereby the molecular length $d$ is altered. Under the assumption that the change in $d$ happens while going through mininum energy configurations nearly adiabatically, we can extract the interaction energy $V(d)$ through a large range of values for $d$ from a single simulation.
Twisted real projective plane boundary conditions are applied as described in Ref.~\cite{Choudhury2022}. 
Here we 
detail the procedure to more accurately fit the interaction energy data 
than in \cite{Choudhury2022}, since small deviations in fitting result in large deviations of the derivative and hence in Fig.~\ref{omega_comparisonvsgpe}. 
We plot the raw data for the interaction energy $V(d)$ from imaginary time evolution (orange dots, only representative data point are shown) together with the parametrization used for the model dynamics. The parameterization is a composite using three different procedures.

\begin{figure}[H]
    \centering
    \includegraphics[width=1\linewidth]{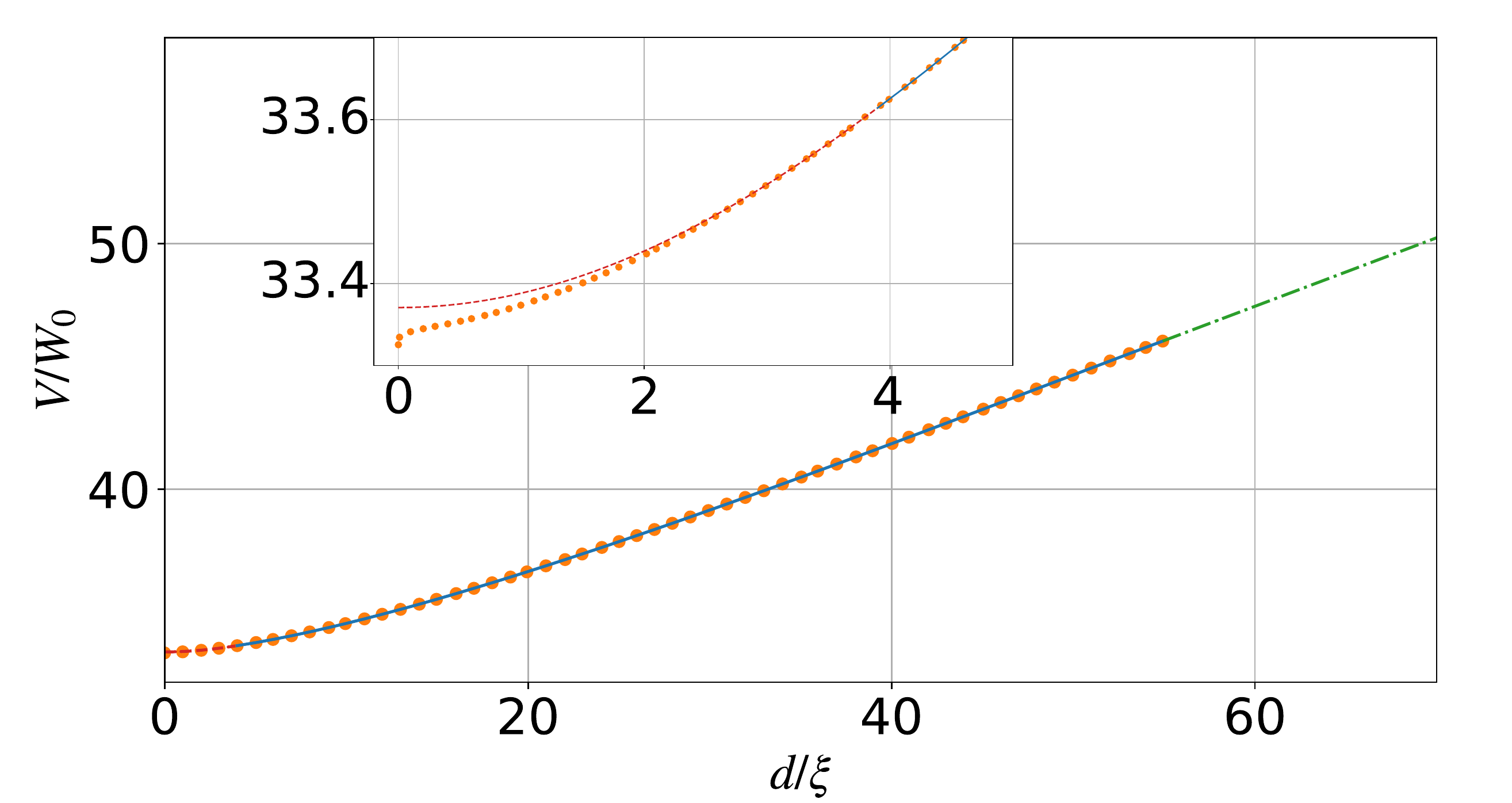}
    \caption{Fitting the interaction energy $V(d)$ as a function of molecular length $d$. 
    The orange dots show data from imaginary time evolution of the GPE from Ref.~\cite{Choudhury2022} (only some representative data points are shown here). The inset is a closeup showing data for small values of $d$. The lines show the parameterization of $V(d)$ used for the model dynamics discussed in this paper consisting of three different segments.
    The blue solid line shows a third order spline interpolation of the numerical data for $4\xi <d< 55\xi$.  
    The green dot-dashed line is a linear extrapolation for large $d > 55\xi$. The red dashed line shows
    the quartic extrapolation of Eq.~\eqref{eq:quartic_extrap} for $d < 4\xi$.
    Parameters are $\nu = 2 \times 10^{-3}\mu$, $g_{12}=0$. $W_0=\hbar^2(\mu+\nu)/m(g+g_{12}) = (\mu+\nu)^2\xi^2/(g+g_{12})$ and an arbitrary offset was added to the energy. 
    }
    \label{vint_fitting}
\end{figure}

The imaginary-time simulation was seeded with phase-imprinted component vortices at an initial distance of $d = 60\xi$ and then evolved to reduce the molecular distance $d$ down to values much smaller than a healing length. We disregard data with $d>55\xi$ where the domain wall is formed and imaginary-time evolution is not adiabatic. The raw data in the interval $3\xi < d < 55\xi$ is considered reliable, were we use a third order spline intepolation of the numerical data in order to obtain a continuous representation for $V(d)$. This is shown as a full blue line in Fig.~\ref{vint_fitting}. Since we expect the interaction energy to be linear at length scales much larger than $\lJ=11.19 \xi$, we perform a linear extrapolation for $d>55\xi$ using the last three  spline points. The linear extrapolation is shown as a dash-dotted green line.

The inset in Fig.~\ref{vint_fitting} shows a closeup for small $d<5\xi$. In this regime the imaginary-time evolution in $\tau$ decreased the molecular length $d$ increasingly slowly while still reducing the energy, presumably by making subtle adjustments to the phase at large range. We thus consider the increasingly sharp drop of the energy near $d=0$ in the imaginary time data an artefact.
Instead we expect the true interaction energy to be an analytic function of the component vortex coordinates and an even function of $d$. Thus, it can be written as a power series in even powers of $d$. We thus extrapolate the interaction energy with a fourth order polynomial for $d<4\xi$
\begin{align} \label{eq:quartic_extrap}
    V(d)=p_1+p_2 d^2 + p_3 d^4 .
\end{align}
The fitting parameters $p_1=33.35 W_0$, $p_2=0.0239 W_0/\xi^2$ and $p_3=-0.000714 W_0/\xi^4$ are obtained from a least-squares fit of the imaginary-time evolution data in the interval $3\xi <d<5\xi$. The quadratic extrapolation is shown as a dashed red line.

In our numerical simulations we have used dimensionless units where $1 = \tilde\mu+\tilde\nu = \tilde\xi = \tilde g + \tilde{g}_{12} = \tilde{n}_0$. The correct unit for the energy functional is thus $W_0=\hbar^2(\mu+\nu)/m(g+g_{12}) = (\mu+\nu)^2\xi^2/(g+g_{12})$.

\section{General integrals of the charge distribution model}
\label{app:dist}
The velocity of the vortex at $\mathbf{R}_1$ in the distributed vorticity model in any arbitrary position of the vortex molecule
is given by Eq.~\eqref{eq:v_dv} as,
\begin{align}
    \mathbf{V}_1^\mathrm{dv}(R_1) &= \kappa \int_0^1 \mathbf{u}_\mathrm{im}(\mathbf{R}_1; {(1-t)\mathbf{R}_1 + t \mathbf{R}_2})  \mathrm{d}t, \\
    & =\frac{\hbar \bar{\kappa}}{m}\hat{z}\times\left[ \mathbf{R}_1 I_1 -    \mathbf{R}_2 I_2 \right]  ,   
\end{align}
where $I_1$ and $I_2$ are 
\begin{align}
    I_1= \int_0^1 \frac{\left(1-\frac{L^2}{R_i^2}\right)+t\frac{L^2}{R_i^2}}{|\mathbf{R_1}\left(1-\frac{L^2}{R_i^2}\right)-\frac{L^2}{R_i^2}t \mathbf{d}|^2} dt, \\
    I_2=\int_0^1 \frac{\frac{L^2}{R_i^2}t}{|\mathbf{R_1}\left(1-\frac{L^2}{R_i^2}\right)-\frac{L^2}{R_i^2}t \mathbf{d}|^2} dt, 
\end{align} 
with $\mathbf{d}=\mathbf{R}_2-\mathbf{R}_1$ and $R_i^2=|\mathbf{R}_1 + t \mathbf{d}|^2$. The scalar integrals come out as,
\begin{align}
    I_1=&\frac{1}{d R_1^3}\left(d R_1 + L^2 \left\{ \biggl[ \arctan(\cot\theta) - \right.\right.  \nonumber \\
   & \left. \arctan\left(\cot\theta + \frac{d R_1 \cosec\theta}{ R_1^2-L^2}\right)\right] \cos2\theta \cosec\theta     \nonumber \\ 
     & \left. \left.   +  \cos\theta\ln\frac{d^2 R_1^2 + (L^2 - R_1^2)^2 + 2 d R_1 (R_1^2-L^2 ) \cos\theta}{(L^2 - R_1^2)^2}\right\}\right) ,  \\
      I_2=&\frac{R^2}{2 d^2 R_1^2} \left\{ 2 \left[\arctan(\cot\theta) - \arctan\biggl(\cot\theta  \right.\right.  \nonumber \\ 
    & \left. \left. + \frac{d R_1 \cosec\theta}{-L^2 + R_1^2}\right)\right] \cot\theta  \nonumber\\ 
      & \left.    + \ln\frac{d^2 R_1^2 + (L^2 - R_1^2)^2 +2 d R_1 (R_1^2-L^2  ) \cos\theta}{(L^2 - R_1^2)^2} \right\} 
      \label{integrals_explicit2} 
\end{align}
with $\theta$ being the angle between $\mathbf{R}_1$ and $\mathbf{d}$. $I_1$ has a removable
singularity at $\theta = 0$.  
\bibliography{Solitons,Books} 
\end{document}